\documentclass[twocolumn,noshowpacs,preprintnumbers,amsmath,amssymb]{revtex4}

\usepackage{verbatim}
\usepackage{wasysym}

\usepackage{graphicx}
\usepackage{dcolumn}
\usepackage{bm}
\usepackage{wrapfig}

\begin{document}

\title{Ultra-thin PbS sheets by two-dimensional oriented attachment}

\author{Constanze Schliehe}
\affiliation{Institute of Physical Chemistry, University of Hamburg, 20146 Hamburg, Germany}
\author{Beatriz H. Juarez}
\affiliation{IMDEA Nanoscience, 28049 Madrid, Spain}
\author{Marie Pelletier}
\author{Sebastian Jander}
\author{Denis Greshnykh}
\author{Mona Nagel}
\author{Andreas Meyer}
\author{Stephan Foerster}
\author{Andreas Kornowski}
\author{Christian Klinke}
\author{Horst Weller}
\email{weller@chemie.uni-hamburg.de}
\affiliation{Institute of Physical Chemistry, University of Hamburg, 20146 Hamburg, Germany}

\begin{abstract} 

Controlling anisotropy is a key concept to generate complex functionality in advanced materials. For this, oriented attachment of nanocrystal building blocks, a self assembly of particles into larger single crystalline objects, is one of the most promising approaches in nanotechnology. We report here the 2D oriented attachment of PbS nanocrystals into ultra-thin single crystal sheets with dimensions on the micrometer scale. We found that this process is initiated by co-solvents which alter nucleation and growth rates during the primary nanocrystal formation and finally driven by dense packing of oleic acid ligands on $\left\{ 100 \right\}$ facets of PbS. The obtained nanosheets can be readily integrated in a photo-detector device without further treatment.

\end{abstract}

\maketitle

Controlled assembly leading to anisotropic nanostructures poses a conceptual challenge in materials research. Penn and Banfield \cite{1,2} described crystal growth, in which oxide nanoparticles coalesce in well defined crystalline orientations. Their method of oriented attachment of nanocrystals is now one of the most favorable techniques to grow linear or zig-zag-type one-dimensional nanostructures. In addition to strong size quantization effects occurring in these structures, their big advantage is solution processability making them attractive candidates for optoelectronic and thermoelectric applications in low-cost integrated systems. One-dimensional assemblies of oriented attachment have been reported, and in most cases the anisotropy during self-assembly is caused by crystal planes with preferred reactivity and dipole moments in the crystallites. Systems with cubic crystal symmetry, however, like PbS and PbSe, where beautiful one-dimensional oriented attachment occurs, are somewhat more difficult to explain. Oriented attachment, in this case, should result in three-dimensional networks rather than one-dimensional structures. The common explanation assumes that despite the strict monodispersity of the samples inhomogeneities in the chemical composition of surface planes exist and result in dipole moments within the nanocrystals. On the other hand organic ligand molecules play a crucial role in such processes by capping nanoparticle surfaces selectively and may hinder, modify, or trigger an oriented attachment \cite{3}. In this work we show that the formation of ordered and densely packed ligand surface layers of oleic acid on $\left\{ 100 \right\}$ PbS surfaces can drive the normally isotropic crystal growths into a two-dimensional oriented attachment of nanocrystals. Hereby the presence of chlorine containing co-solvents during the initial nucleation and growth process of the nanocrystals plays a prominent role.

\begin{figure}[htbp]
  \centering
  \includegraphics[width=0.45\textwidth]{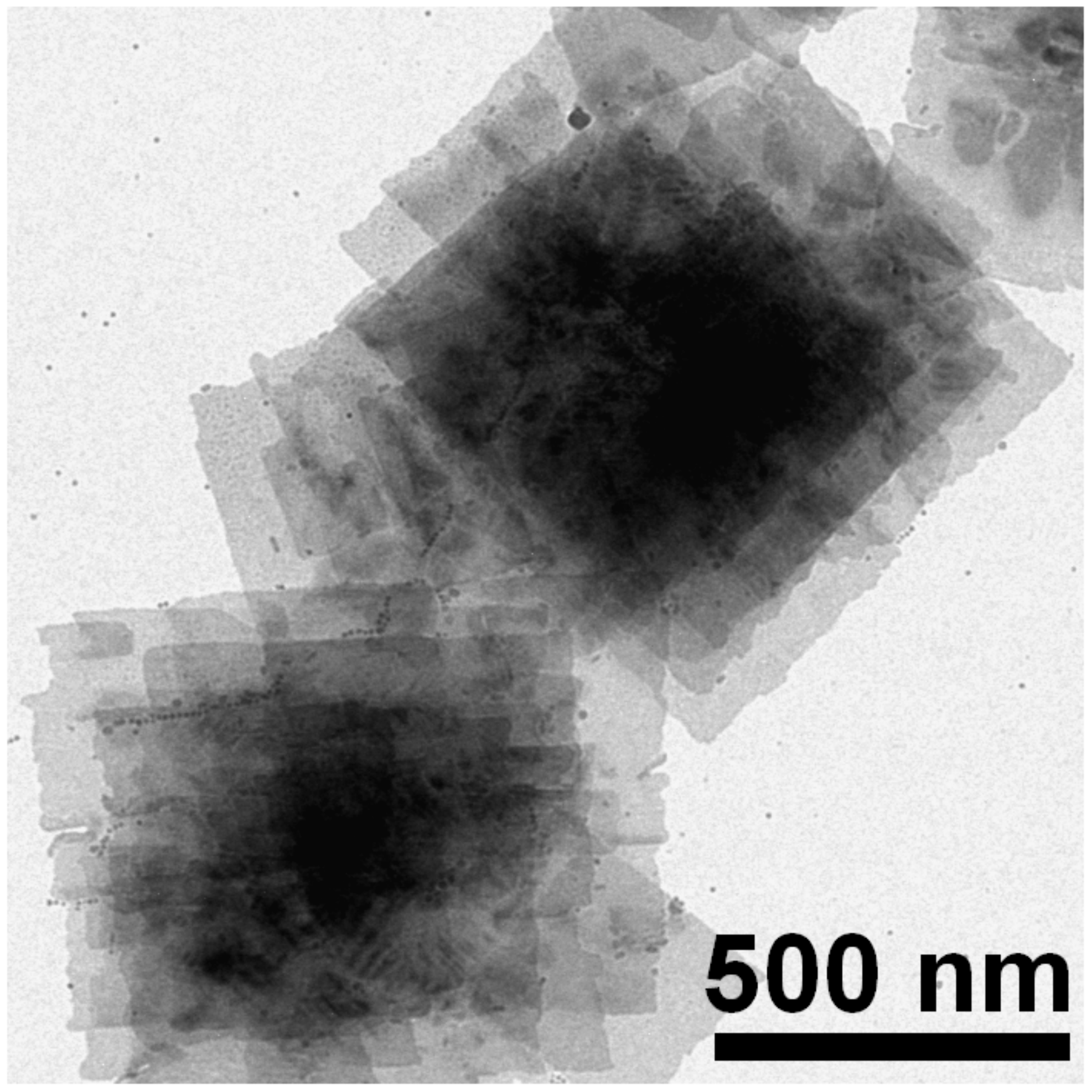}
  \caption{\textit{TEM image of stacked PbS nanosheets using 1,1,2-trichloroethane TCE.}}
\end{figure}

Syntheses of lead chalcogenide nanoparticles \cite{4,5,6,7} can lead to a large variety of particle shapes through slight changes of the reaction conditions. It has been reported how small traces of organo-phosphine compounds may alter the course of the reactions, which underlines their complexity \cite{8}. For other systems, like CdS nanoparticles, shape control has been reported by adding of small amounts of HCl \cite{9} or chlorine containing solvents like 1,2-dichloroethane (DCE) for example in the case of CdSe nanorods attached to carbon nanotubes \cite{10}. Reported strategies to generate 2D nanostructures are based on the use of lamellar-like templates \cite{11} or thin superstructures by assembly \cite{12,13,14}.

\begin{figure}[htbp]
  \centering
  \includegraphics[width=0.45\textwidth]{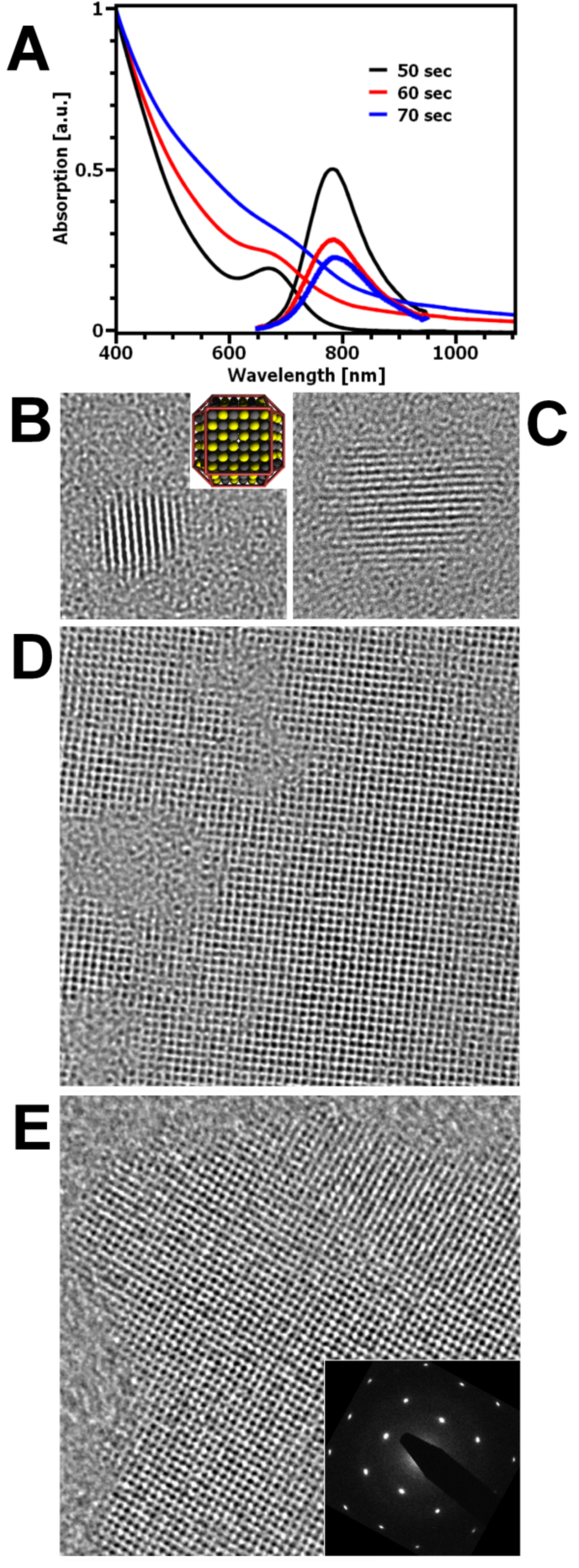}
  \caption{\textit{(A) Absorption and emission spectra of the nanoparticles with reaction time. The absorption spectra were normalized to the absorption at 400 nm, and the emission spectra to the maximum of each curve. (B) TEM image of an ultra-small PbS nanoparticle at early stages of the reaction. Inset: Model structure of a truncated cuboctahedron. (C) First step of attachment. (D) Later the structures merge with some holes due to non-ideal attachment. (E) At the end of the synthesis quasi continuous sheets are formed. Inset: Electron diffraction pattern of the PbS sheets.}}
\end{figure}

Our approach is based on a standard synthetic procedure to generate nearly spherical PbS nanoparticles \cite{15}. After 3 minutes the resulting dot-like particles have a mean diameter of around 5 nm. In contrast, ultrathin PbS nanosheets as depicted in Fig. 1, are formed in the presence of chlorine containing compounds like 1,2-dichloroethane or similar linear chloroalkanes. The nanosheets have lateral dimensions of several hundred nanometers and stack showing Moire patterns caused by the interference between the crystalline lattices of the individual sheets. The formation of the sheets occurs within the first one to three minutes of the reaction. To investigate the mechanism of formation we followed the temporal evolution of the sheets by optical spectroscopy and HRTEM. For practical reasons we slowed down the sheet formation by reducing the reaction temperature immediately after the thioacetamide injection. Fig. 2A shows the absorption and emission spectra at various stages of the reaction. Within the first 50 seconds the solution turned red-brown and clearly structured spectra evolved with an absorption maximum at 675 nm and a narrow emission band at 790 nm. These features indicate small isolated quantum dots of 2.8 $\pm$ 0.5 nm, which can clearly be identified by TEM. As the reaction evolved the fluorescence band of the quantum dots decreased and an unstructured absorption of sheets arose, partially due to the occurrence of turbidity. The spectral position of the original bands, however, and thus the size of the particles, does not change during this process proving that sheets are formed directly from the nanoparticles by oriented attachment. The growth of either ultrathin sheets or larger particles (in the absence of DCE) is determined by the behavior of the PbS nanocrystals originally formed. 

Nanocrystals as small as 2.8 nm could not be isolated if DCE (or a similar chlorine containing compound) was not added to the reaction mixture. A TEM sample taken during the rapid process of oriented attachment shows various stages of sheet formation. The original particles (Fig. 2B) can be recognized together with fused aggregates (Fig. 2C) as well as porous structures similar to the one shown in Fig. 2D. In Fig. 2B, where an individual particle in a slightly tilted $\left[ 100 \right]$ orientation is depicted, an angle of 135$^{\circ}$ at the crystalline edges can be identified. This is taken as an indication for the presence of $\left\{ 110 \right\}$ surface planes. A reasonable model for such a small crystallite is shown in the inset of this figure. A close inspection of Fig. 2C identifies this structure as an aggregate of three nanocrystals fused in 2 dimensions. Even in the porous framework structure of Fig. 2D the size and shape of the original particles as well as their crystalline orientation is visible. Uniform nanosheets, however, can be obtained several minutes after the sulfur precursor injection (Fig. 2E). The HRTEM image of a sheet region depicted in Fig. 2E shows a single crystalline structure with crossed (200) lattice planes displaying an angle of 45$^{\circ}$ relative to the sheet edges. This indicates a sheet growth parallel to the $<$110$>$ axes. Although most of their edges in $<$110$>$ directions are atomically flat over large distances, some regions clearly show a zig-zag pattern with exposed $\left\{ 100 \right\}$ facets with dimensions corresponding to those of the original quantum dots. Selected area electron diffraction (SAED) pattern of individual sheets confirm this observation showing a monocrystalline galena structure in $\left[ 100 \right]$ orientation with pronounced (200) and (220) reflexes. The X-ray diffraction pattern shows that the HRTEM characterization of individual sheets is representative of the entire sample.

\begin{figure}[htbp]
  \centering
  \includegraphics[width=0.45\textwidth]{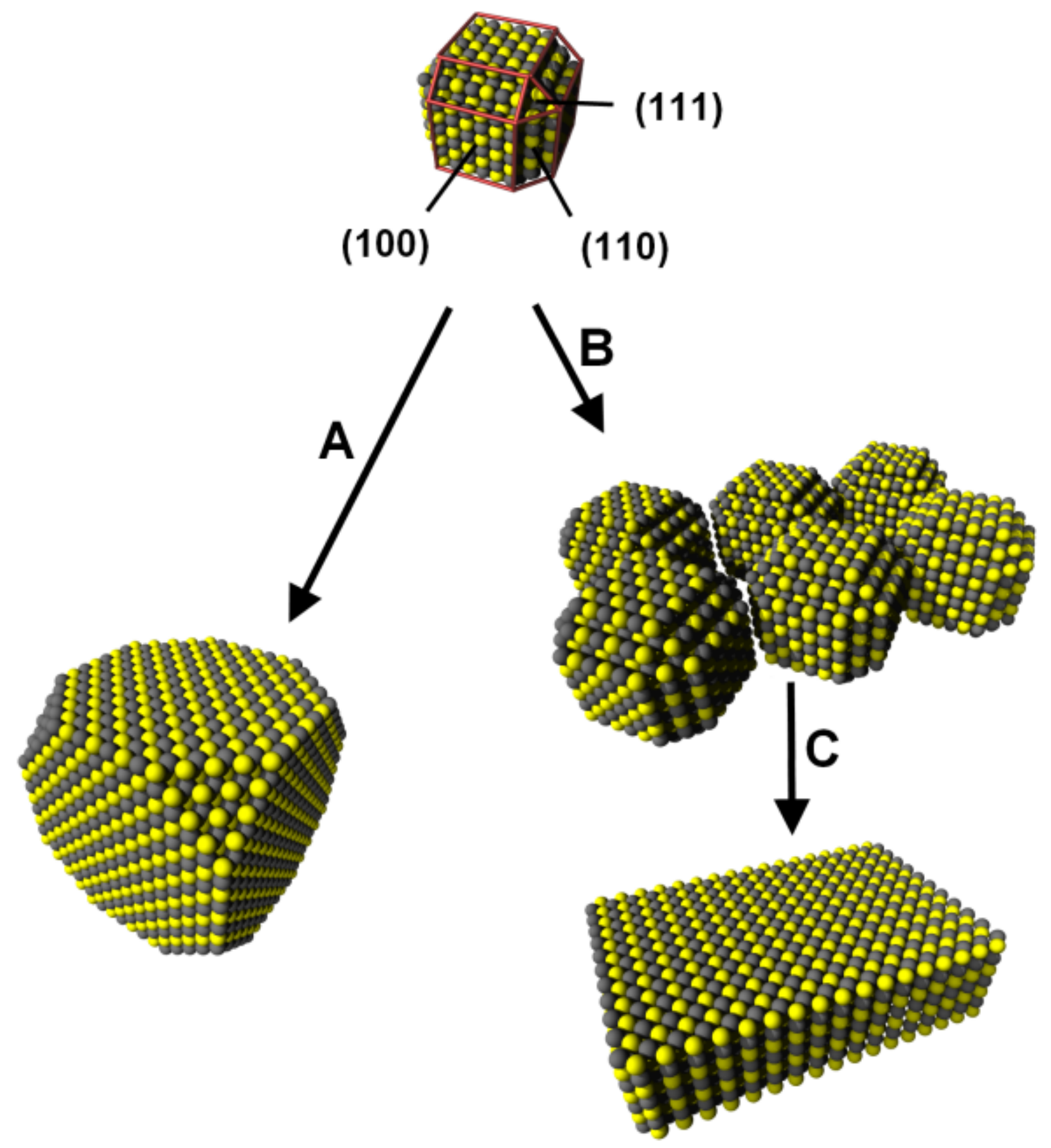}
  \caption{\textit{Schematic illustration of large particle (A) and sheet formation (B, C) from small PbS quantum dots.}}
\end{figure}

Murray and co-workers found in the crystalline equivalent PbSe and PbTe systems that truncated cubes with sizes of roughly 10 nm evolve from approximately 5 nm cuboctahedrons with 6 $\left\{ 100 \right\}$ and 8 $\left\{ 111 \right\}$ facets. They explained this by a faster growth perpendicular to the $\left\{ 111 \right\}$ facets \cite{16,17}. These results also suggest that $\left\{ 110 \right\}$ facets are highly reactive and therefore are preferentially consumed during the nanoparticle growth. The structure of PbS nanocrystals which merge into PbS sheets is shown in Fig. 3. The shape is that of a truncated cuboctahedron with 6 $\left\{ 100 \right\}$, 8 $\left\{ 111 \right\}$ and 12 $\left\{ 110 \right\}$ facets. Upon crystal growth the particles try to minimize the most energetically unfavorable surface facets by a fast growth perpendicular to the respective facet. In the case of PbS particles bigger than 10 nm \cite{18} this results in truncated cubes with mainly $\left\{ 100 \right\}$ surface planes (path A in Fig. 3). We do not observe the formation of PbS sheets in the absence of DCE (or a similar chloride compound), whereas in the presence of chloride compounds, which are known to act as lead complexing agents \cite{19}, the kinetics of nucleation and growth are altered, leading to approximately 3 nm particles (as shown in Fig. 2A) which should still exhibit the reactive $\left\{ 110 \right\}$ surfaces. Since Ostwald ripening plays only a negligible role during PbS nanoparticle formation, these small crystallites stabilize by oriented attachment via the $\left\{ 110 \right\}$ facets (path B in Fig. 3). The expected egg tray like structure is unlikely to be stable and will undergo a surface reconstruction (path C in Fig. 3) resulting in a flat sheet (Fig. 2E). The ensuing sheet should be thinner than the original building blocks by up to 30 percent. This reconstruction, understood as the diffusion and accommodation of the atoms under favorable thermodynamic conditions (100$^{\circ}$C) also contributes to a decrease in the porosity of the sheets. We performed small-angle x-ray scattering (SAXS) of the stacked PbS layers (Fig. 4). The sample was positioned in a coplanar configuration with the SAXS beam in order to obtain information about the sheet thickness and orientational distribution. A highly anisotropic dumbell-type scattering pattern characteristic for sheets oriented nearly parallel to the q$_{x}$-direction can be seen in the upper inset of Fig 4. The best simulation fit was calculated assuming a regular stacking of disks with a long period of 5.8 nm and PbS sheet thickness of 2.2 nm. The resultant distance between the sheets of 3.6 nm is attributed to an oleic acid bilayer. This behavior strongly supports the notion of an interplanar crystalline oleic acid bilayer, not understandable by randomly coiled and interpenetrating chains. The formation of the stacked layers occurs most probably by an attachment of quantum dots or small sheet structures on top of already formed nanosheets, which than acts as a nucleus for further 2D attachment. The stacked structure can also be observed by AFM. 

\begin{figure}[htbp]
  \centering
  \includegraphics[width=0.45\textwidth]{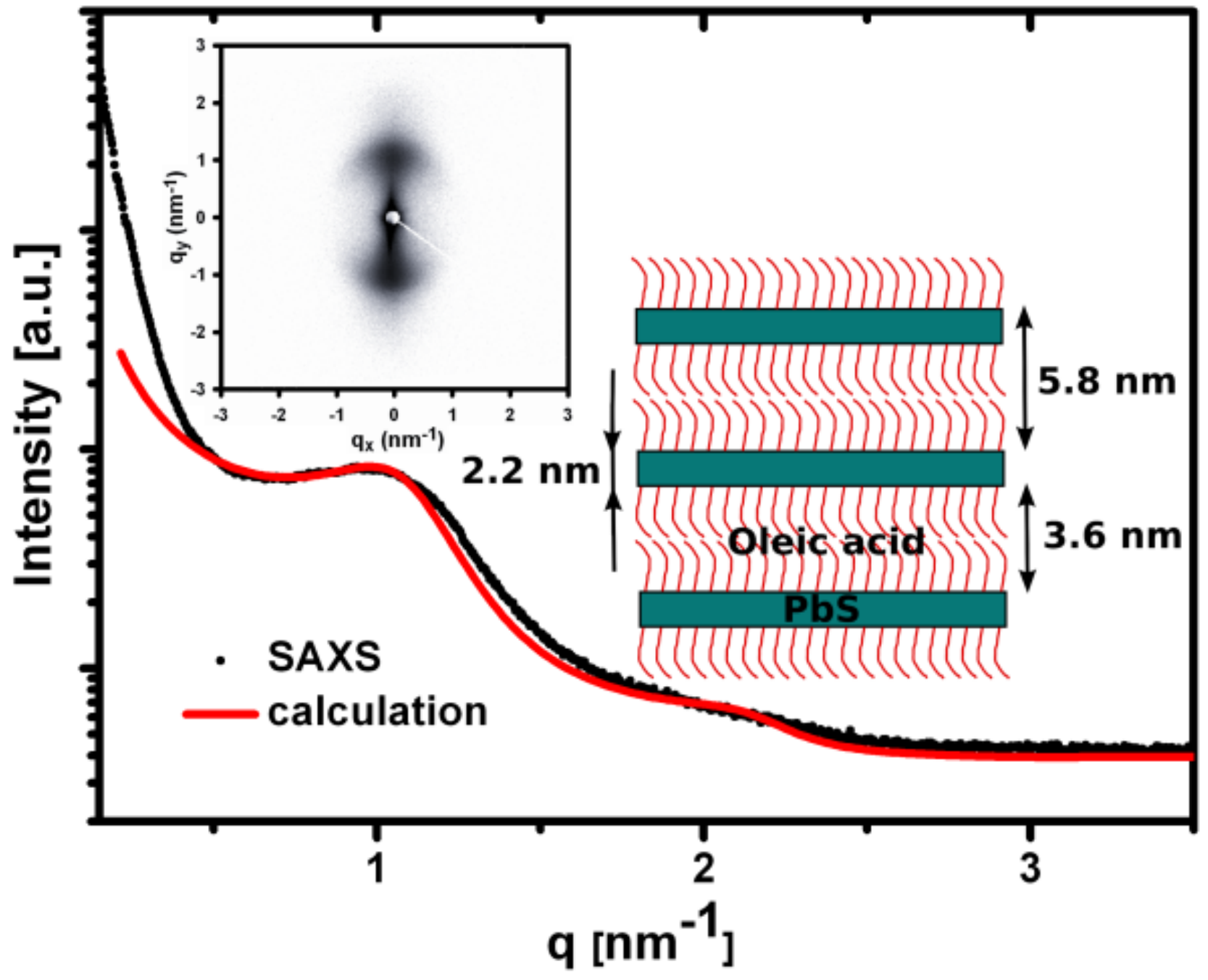}
  \caption{\textit{Cross section of the scattering intensity along the q$_{y}$ axis (black points) and simulated diffraction curve. The scattering image is depicted in the inset. The result could be modeled with assumed disks of 2.2 nm thickness and a spacing ligand of 3.6 nm.}}
\end{figure}

\begin{figure}[htbp]
  \centering
  \includegraphics[width=0.45\textwidth]{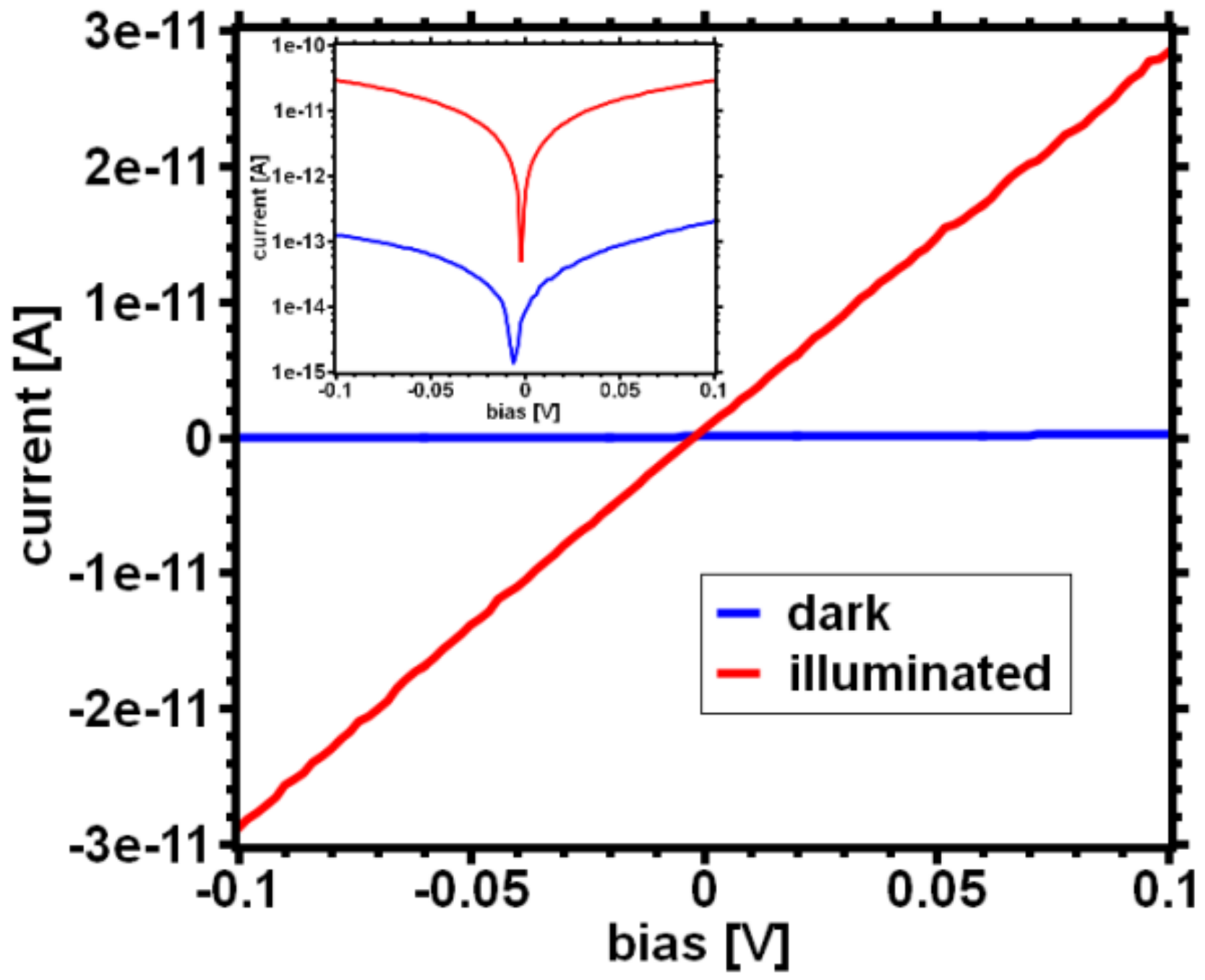}
  \caption{\textit{I-V curves of PbS sheets with and without illumination with a green laser. The thin film of stacks of PbS sheets cover a detection area of 0.5 $\mu$m x 6 $\mu$m between two gold electrodes.}}
\end{figure}

One question remains, however, as to why oriented attachment leads to 2D instead of a 3D structure. Our results strongly support that the assembly of oleic acid in a highly ordered monolayer on the $\left\{ 100 \right\}$ surfaces plays a key role in the formation and further stacking of PbS sheets. Oleic acid is well known to assemble vertically on various substrates and to form bilayer structures \cite{20,21,22}. The thickness of a monolayer was determined from single crystal data and ranges from 1.7 to 2.4 nm in the reported crystal structures. The latter mainly varies by the conformation next to the double bond and the relative orientation of the oleic acids within one monolayer. These reported values fit well with the observed interplanar distance of 3.6 nm in stacked sheets. The formation of a dense and highly ordered oleic acid layer covering the PbS nanosheets also allows the system to bind a maximum number of oleic acid molecules. The resulting release of adsorption enthalpy combined with the stabilization by inter-alkyl van der Waals forces obviously overcompensates the decrease in entropy and results in a net release of Gibbs energy. The 2D oriented attachment and the reconstruction from the egg tray structure to flat sheets would, from this point of view, not only be driven by minimizing high energy surfaces, but also by structural changes caused by the oleic acid molecules. Whereas one expects the ligands to exist in a randomly coiled state on the highly curved surface of small particles, in fact they appear in a highly ordered phase on the flat $\left\{ 100 \right\}$ planes. We believe that similar effects may hold in general in shape control, self-assembly, and templated growth of nanocrystals. For example, ligand driven self-assembly of nanocrystals into aggregated wires (1D) and vesicles (curved 2D) was recently observed for amphiphilic CdSe quantum dots \cite{23}.
 
The thin film structure of the sheets should decisively affect the quantum confinement of photogenerated carriers. In accord with the reported dimensions, no confinement is expected in lateral direction while the structures should be strongly confined vertically. Differences in layer thickness and confinement in one dimension may strongly influence the optical features. Correlated confocal microscopy and TEM inspections confirms this assumption showing an emission signal coming from two different individual stacks of PbS sheets at a wavelength of about 720 nm. 

Lead chalcogenide nanoparticle arrays are currently in the spotlight for photodetector fabrication. Typical 2D assemblies require physical or chemical treatments to remove the insulating organic shell capping \cite{24,25,26}. The lack of ligands in the in-plane dimensions of raw PbS 2D sheets produces intrinsically outstanding photoconductive properties, as shown in Fig. 5. PbS sheets bridging two gold electrode on a SiO$_{2}$ substrate exhibit low dark conductance between $\pm$0.1 V without substantial hysteresis. Upon illumination with a 532 nm laser, however, the conductance increases by more than two orders of magnitude at an illumination intensity of 2.0 mW/cm$^{2}$. This yields an efficiency value (actual photocurrent divided by the theoretical max. current due to photon impact) of about 1.1 at 0.1 V (corresponds to a responsivity of 0.472 A/W). Taking into account that the number of absorbed photons is only a small fraction of the incident, this value shows that the increase in current cannot be due to harvesting of photogenerated charge carriers only. The filling of trap states is most probably responsible for the current gain. In contrast to currently investigated devices made from ligand stabilized PbS nanocrystals \cite{27,28} the ready-made, non-treated nanosheets reported here are remarkably conductive due to missing in-plane ligands and continuous connection through the monocrystal.

In conclusion we have shown that chlorine containing co-solvents can direct the colloidal synthesis of PbS nanoparticles into two-dimensional nanosheets by oriented attachment. We believe that this is a consequence of the size of the nanoparticles and hence the presence of $\left\{ 110 \right\}$ reactive facets, which disappear as the nanoparticle grow. The presence of DCE or similar chlorine-containing compounds yields small nanoparticles of around 3 nm in diameter, exposing the highly reactive $\left\{ 110 \right\}$ facets so as to trigger an oriented attachment process. Both the oriented attachment and the strong crystalline correlation between the nanosheets in stacked aggregates are most likey driven by the formation of a highly ordered oleic acid bilayer between the PbS nanosheets. The absence of ligands in the in-plane direction of the ultra thin sheets yields a prominent photoconductivity response in the pristine dried nanomaterial.

\end{document}